\documentclass[%
preprint,
showpacs,
 amsmath,amssymb,
 aps,
 prl,
]{revtex4-1}

\usepackage{cmap}                   
\usepackage[T2A]{fontenc}           
\usepackage[utf8]{inputenc}         
\usepackage[english]{babel} 

\usepackage{graphicx}
\usepackage{dcolumn}
\usepackage{bm}

\usepackage{hyperref}

\newcommand{\id}{{\rm i}\,}

\begin{document}

\makeatletter

\title{Frustrated Heisenberg Antiferromagnets on Cubic Lattices: Magnetic Structures, Exchange Gaps, and Non-Conventional Critical Behaviour}

\author{A.\,N.\,Ignatenko}
\email{ignatenko@imp.uran.ru}
\affiliation{Institute of Metal Physics, Kovalevskaya Str., 18, 620990, Ekaterinburg, Russia}

\author{V.\,Yu.\,Irkhin}
\affiliation{Institute of Metal Physics, Kovalevskaya Str., 18, 620990, Ekaterinburg, Russia}

\date{\today}

\begin{abstract}
We have studied the Heisenberg antiferromagnets characterized by the magnetic structures with the periods being two times larger than the lattice period. We have considered all the types of the Bravais lattices (simple cubic, bcc and fcc) and divided all these antiferromagnets into 7 classes i.e. 3 plus 4 classes denoted with symbols A and B correspondingly. The order parameter characterizing the degeneracies of the magnetic structures is an ordinary Neel vector for A classes and so-called 4-complex for B classes. We have taken into account the fluctuation corrections for these states within the spin-wave and large-N expansions  ($N$ is the number of spin components). Below the Neel temperature $T_{\rm{N}}$ quantum and thermal fluctuations lift the degeneracy making simple one-wave vector collinear structure preferable for all the classes. A satellite of this effect is opening of the exchange gaps at certain wave vectors in the spin wave spectrum (there is an analogous effect for the nonuniform static transverse susceptibility). However, as the temperature approaches $T_{\rm{N}}$, the exchange gaps are closing. We have calculated the critical indices $\eta$ and $\nu$ to order of $1/N$ and found that they differ for A and B classes.
\end{abstract}


\maketitle

Complex exchange interactions (frustrations) can cause the system to demonstrate interesting and unusual behavior. Shender \cite{Shender_1982} studied a garnet antiferromagnet with the magnetic structure consisting of two antiferromagnetic sublattices with Neel vectors which do not interact with each over at the mean-field level. He showed that when fluctuations are taken into account, the interaction between the sublattices appears causing their Neel vectors to order collinearly and opening local exchange gaps in the spin wave spectrum \cite{Shender_1982}.

In this paper we study the classes of antiferromagnets where a similar situation takes place provided that the wave vector of the magnetic structure is not invariant under the lattice symmetry transformations, so that the star of the magnetic wave vector contains several wave vectors. For simplicity we restrict ourselves to the Bravais lattices with cubic symmetry and consider only the magnetic structures with the periods being two times larger than the lattice period. The Fourier expansions of these structures contain only wave vectors $\mathbf{Q}=\mathbf{K}/2$ equal to half of the reciprocal lattice vectors $\mathbf{K}$. For $D$-dimensional Bravais lattice there are $2^D-1$ wave vectors of this form. Combining these wave vectors into the  stars for simple cubic, bcc, and fcc lattices we obtain the table \ref{table_1} with 7 classes.

The wave vectors of the magnetic structures in the Heisenberg model $H=\frac{1}{2}\sum_{ij}J_{ij}\mathbf{S}_{i}\cdot\mathbf{S}_{j}$ on Bravais lattices are determined (neglecting thermal and quantum fluctuations) by the minima of the Fourier transform $J(\mathbf{k})$ of the exchange parameters $J_{ij}$. Since the wave vectors within each of the classes are degenerate, the corresponding magnetic structure has the form of a linear combination
\begin{equation}
\label{eqn_order_parameter}
\mathbf{S}_{i}=\sum_{s=1}^{L}\mathbf{a}_{s}\,e^{\id \mathbf{Q}_{s}\cdot\mathbf{r}_{i}}.
\end{equation}
Due to the relations $\mathbf{S}_{i}^2=1$, real vectors $\mathbf{a}_s$ should satisfy additional constraints $\sum_{s=1}^{L}\mathbf{a}_{s}^2=1$ and $\sum_{\mathbf{Q}_{s}+\mathbf{Q}_{t}\cong\mathbf{W}_{\alpha}} \mathbf{a}_{s}\cdot \mathbf{a}_{t} = 0$, where $M$ vectors $\mathbf{W}_\alpha$ are defined in the table \ref{table_1}. The magnetic structure (\ref{eqn_order_parameter}) for the classes with $L>1$ can be simplified by performing linear transformations mixing vectors $\mathbf{a}_s$. Thus the magnetic structure for the class A$_4$ consists of 4 simple cubic sublattices, each having its own Neel vector corresponding to $(\pi,\pi,\pi)$ staggered order (AF-II structure in the notations of Ref.\cite{Yamamoto_JPSJ_1972}). The magnetic structures for B classes cannot be reduced to independent staggered orders on several sublattices and have different form. For the classes B, B$^{\prime}$ and B$^{\prime\prime}$ the magnetic structure is described by $L=3$ mutually orthogonal vectors $\mathbf{a}_{s}$ of unit total length. Since exponential factors in (\ref{eqn_order_parameter}) take only $\pm 1$ values, magnetic moments $\mathbf{S}_{i}$ in this structure are aligned along 4 directions in spin space, which are not independent. Hence we call this structure 4-{\it complex} (in this terminology, ordinary staggered magnetic structure should be called 1-{\it complex}). The magnetic structure for the remaining class B$_2$ with bcc lattice is built from two independent 4-complexes with the wave vectors $S_I$ (see  table \ref{table_1}), each residing on its own simple cubic sublattice of the bcc lattice.
\begin{table*}[t]      
\caption{\label{table_1} 7 classes of antiferromagnets with the doubly-periodic magnetic structures on cubic Bravais lattices. Here $\mathbf{Q}_s$ are the vectors of the wave vector star, $\mathbf{W}_{\alpha}=\mathbf{Q}_{s}+\mathbf{Q}_{t}$ for $s\ne t$.}
\begin{tabular}{|c|c|c|c|c|c|}
\hline
class               &lattice& $L$ & $\mathbf{Q}_s$, $s=1,\ldots L$                                               & $M$  & $\mathbf{W}_{\alpha}$, $\alpha=1,\ldots M$\\
\hline
A                   & sc    &  1  & $(\pi,\pi,\pi)$                                                              & 0    & --- \\
B                   &       &  3  & $(\pi,0,0),\; (0,\pi,0),\; (0,0,\pi)$                                        & 3    & $S_{\mathrm{I}}$\\
B$^{\prime}$        &       &  3  & $S_{\mathrm{I}}=\{ (0,\pi,\pi),\; (\pi,0,\pi),\; (\pi,\pi,0) \}$             & 3    & $S_{\mathrm{I}}$ \\       
\hline
A$^{\prime}$        & bcc   &  1  & $(2\pi,2\pi,2\pi)$                                                           & 0    & --- \\
B$_2$               &       &  6  & $S_{\mathrm{I}}\cup S_{\mathrm{I}}^{\prime}$ & 7 & $S_{\mathrm{I}}\cup S_{\mathrm{I}}^{\prime}\cup (2\pi,2\pi,2\pi)$ \\
& & & $(S_{\mathrm{I}}^{\prime} =\{ (0,\pi,-\pi),\; (\pi,0,-\pi),\; (\pi,-\pi,0) \})$ & &  \\
\hline
B$^{\prime\prime}$  & fcc   &  3  & $S_{\mathrm{II}}=\{ (2\pi,0,0),\; (0,2\pi,0),\; (0,0,2\pi) \}$               & 3    & $S_{\mathrm{II}}$ \\ 
A$_4$               &       &  4  & $(-\pi,\pi,\pi)$, $(\pi,-\pi,\pi)$, $(\pi,\pi,-\pi)$, $(\pi,\pi,\pi)$        & 3    & $S_{\mathrm{II}}$ \\
\hline
\end{tabular}
\end{table*}

We have calculated the free energy for the general configuration (\ref{eqn_order_parameter}) with at least 3 vectors $\mathbf{a}_s$ being non-zero within the spin-wave theory (SWT) and have shown that, regardless of the particular form of $J(\mathbf{k})$, quantum and thermal fluctuations always make simple collinear configuration $\mathbf{S}_{i}=\mathbf{a}_{s} e^{\id \mathbf{Q}_s\cdot r_{i}}$ with only one $\mathbf{a}_s\ne 0$ more preferable (the details will be published elsewhere). This lifting of the degeneracy by fluctuations provides an example of the ``order from disorder'' phenomenon \cite{Villain_et_al_1980}.

In spite of lifted degeneracy, the SWT spectrum of excitations above collinear state with only $\mathbf{a}_1\ne 0$, $\omega_{0}(\mathbf{k})=S\sqrt{[J(\mathbf{k})-J(\mathbf{Q}_{1})][J(\mathbf{k}+\mathbf{Q}_{1})-J(\mathbf{Q}_{1})]}$, contains excess number of zero modes at wave vectors $\mathbf{k}=\mathbf{Q}_s$ and $\mathbf{k}=\mathbf{Q}_s-\mathbf{Q}_1$ for the classes with $L>1$ (an ordinary collinear antiferromagnet with $L=1$ has only two Goldstone zero modes). For the classes B$^\prime$, B$^{\prime\prime}$, and B$_2$ an overlap of zeros of two brackets under the square root in $\omega_0(\mathbf{k})$ occurs, which provides $q^2$ dispersion of the corresponding modes. We have checked that the first fluctuation correction to the spin-wave spectrum opens local exchange gaps for all wave vectors $\mathbf{k}=\mathbf{Q}_s-\mathbf{Q}_1$, $\mathbf{Q}_s$, except for $\mathbf{k}=0$, $\mathbf{Q}_1$, and removes the $q^2$ dispersion. Note that these exchange gaps depend on temperature and have crucial influence on the thermodynamics of the system. Most interesting is the behavior of the exchange gaps near $T_\mathrm{N}$, where all spin-wave type theories fail. 

A closely related phenomenon is  opening of the exchange ``gaps'' in the quasimomentum dependence of the inverse transverse static susceptibility $\chi_{t}(\mathbf{k})^{-1}$. In the following we consider this within the $1/N$-expansion for the classical Heisenberg model. In this expansion three-dimensional classical spins are replaced by $N$-component vectors and rescaling of the inverse temperature $\beta=N \bar{\beta}$ is performed. The formal structures of $1/N$-expansions for the Heisenberg, nonlinear sigma, and $\phi^4$ models are similar (for the latter see, e.g., \cite{Ma_1976}). To first order in $1/N$, the inverse susceptibility in the paramagnetic phase and its transverse part in the ordered phase have the form
\begin{equation}
\label{eqn_chi_1_N}
\chi_{\mathrm{para},t}(\mathbf{k})^{-1}=J(\mathbf{k})-J(\mathbf{Q}_1)+\Delta_0^2+\Sigma_{1}(\mathbf{k})+\Sigma_{1}^\prime,
\end{equation}
where $\Delta_0$ is proportional to inverse correlation length $\xi_0^{-1}$ at $N=\infty$,
\begin{equation}
\label{eqn_sigma_1}
\Sigma_{1}(\mathbf{k})=\frac{1}{\mathcal{N}}\sum_{\mathbf{q}}\frac{2}{N}\frac{G(\mathbf{k}+\mathbf{q})-G(\mathbf{Q}_1+\mathbf{q})}{\Pi(\mathbf{q})+2\bar{\beta}\sigma^2 G(\mathbf{Q}_1+\mathbf{q})},\qquad G(\mathbf{k})=\frac{1}{J(\mathbf{k})-J(\mathbf{Q}_1)+\Delta_0^2}
\end{equation}
(here we suppose that in the ordered state $\mathbf{a}_1\ne 0$), in the ordered phase $\Sigma_{1}^\prime=0$, and in the paramagnetic phase $\Sigma_{1}$ can be found from the sum rule for the susceptibility, which gives $\Sigma_{1}^\prime=-\frac{1}{\mathcal{N}}\sum_{\mathbf{k}}G(\mathbf{k})^2\Sigma_{1}(\mathbf{k})/\Pi(0)$.  In the denominator of  eq. (\ref{eqn_sigma_1}), $\sigma$ means the staggered magnetization at $N=\infty$, and $\Pi(\mathbf{q})=\frac{1}{\mathcal{N}}\sum_{\mathbf{k}}G(\mathbf{k})G(\mathbf{k}+\mathbf{q})$.

It can be seen from eqs. (\ref{eqn_chi_1_N}) and (\ref{eqn_sigma_1}) that in the ordered phase ($\sigma>0$) the correction $\Sigma_1(\mathbf{k})$ removes the poles of $\chi_t(\mathbf{k})$ at all $\mathbf{k}=\mathbf{Q}_{s}$, except for $\mathbf{k}=\mathbf{Q}_{1}$. This happens solely because of the second term $\propto\sigma^2$ in the denominator of eq. (\ref{eqn_sigma_1}) violates the symmetry under the lattice transformations. Correspondingly, at the critical point $\bar{\beta}=\bar{\beta}_c$, where $\sigma=0$, the symmetry is restored and the susceptibility $\chi(\mathbf{k})$ has the singularities (but not poles) at all $\mathbf{Q}_{s}$. In the paramagnetic phase these singularities are replaced by finite peaks of the identical form.

Since the susceptibility $\chi(\mathbf{k})$ at the critical temperature contains $L$ singular points, it is natural to expect that the critical behavior for the classes with $L>1$ can differ from the standard one. Generally, the susceptibility is an anisotropic function of $\mathbf{k}$ near each $\mathbf{Q}_s$. However, since this anisotropy is irrelevant to the critical behavior, one can assume the susceptibility being spherically symmetrical near each point $\mathbf{Q}_s$. Moreover, by the same reasons one can perform the substitution
\begin{equation}
\label{eqn_G_approx}
G(\mathbf{k})=\frac{1}{J(\mathbf{k})-J(\mathbf{Q}_1)+\Delta_0^2}\to\sum_s \frac{\theta(\Lambda-|\mathbf{k}-\mathbf{Q}_s|)}{(\mathbf{k}-\mathbf{Q}_s)^2+\xi_0^{-2}},
\end{equation}
where we have introduced the cutoff $\Lambda\ll \pi/a$ for the quasimomentums keeping only small spheres near each $\mathbf{Q}_s$.

We start from the calculation of the index $\eta$ defined by $\chi(\mathbf{k})\sim|\mathbf{k}-\mathbf{Q}_s|^{-2+\eta}$ at the critical point near each $\mathbf{Q}_s$. Performing the substitution (\ref{eqn_G_approx}) in the definition of $\Pi(\mathbf{k})$ one obtains 
\begin{equation}
\label{eqn_1_Pi_approx}
\frac{1}{\Pi(\mathbf{q})}\to \sum_{\alpha}\frac{8|\mathbf{q}-\mathbf{W}_{\alpha}|}{L_{\alpha}}\,\theta(\Lambda-|\mathbf{q}-\mathbf{W}_{\alpha}|),
\end{equation}
for all $\bar{\beta}\ge \bar{\beta}_c$. Here  $\alpha=0,\ldots M$ counts for the vectors $\mathbf{W}_{\alpha}$ defined in the table \ref{table_1} with an addition $\mathbf{W}_{\alpha=0}=\mathbf{0}$, and $L_{\alpha}$ equals to the number of ordered pairs $(s,t)$ satisfying $\mathbf{Q}_s+\mathbf{Q}_t \cong \mathbf{W}_{\alpha}$. Substituting eqs. (\ref{eqn_G_approx}) and (\ref{eqn_1_Pi_approx}) into eq. (\ref{eqn_sigma_1}) and calculating the integral one obtains 
\begin{equation}
\Sigma_{1}(\mathbf{k})\to -\sum_{s, \alpha}\frac{8}{3\pi^2 N}\frac{k_{s,\alpha}^2 \ln k_{s,\alpha}}{L_{\alpha}}\;\theta(\Lambda-k_{s,\alpha}), \qquad \mathbf{k}_{s,\alpha}=\mathbf{k}-\mathbf{Q}_s-\mathbf{W}_{\alpha}.
\end{equation}
Considering here, e.g., $\mathbf{k}\approx \mathbf{Q}_1$ and reducing  similar terms for each class of the table \ref{table_1} separately, one obtains  
\begin{equation}
\Sigma_{1}(\mathbf{k}\approx \mathbf{Q}_1)\to -\frac{8 r}{3\pi^2 N} k_{1,0}^2\ln k_{1,0},
\end{equation}
where the coefficient $r=1$ and $r=4/3$ for the A and B classes correspondingly. Substituting this result into eq. (\ref{eqn_chi_1_N}) and rewriting it in the exponential form we get the result $\eta=8r/3\pi^2 N$ (see the second column of the table \ref{table_2}).

The index $\nu$ is determined from the correlation length, $\xi\sim (\bar{\beta}_c-\bar{\beta})^{-\nu}$. Define $\xi^{-1}$ as an imaginary part of the pole of the function $\chi(\mathbf{k})$ near arbitrary wave vector among $\mathbf{Q}_s$. Then to the first order in $1/N$
\begin{equation}
\label{eqn_corr_to_xi0}
\xi^{-2}-\xi_0^{-2}\approx\Sigma_1(\mathbf{Q}_s+\id \xi_0^{-1})+\Sigma_1^{\prime}=\frac{1}{\mathcal{N}}\sum_{\mathbf{q}}\frac{2}{N\Pi(\mathbf{q})}\left(G(\mathbf{Q}_s+\id \xi_0^{-1}+\mathbf{q})+\frac{1}{2\Pi(0)}\frac{\partial\Pi(\mathbf{q})}{\partial\xi_0^{-2}}\right).
\end{equation}
Performing the same steps as in the calculation of the index $\eta$ and using 
\begin{equation}
\Pi(\mathbf{q})\to \sum_{\alpha} L_{\alpha}\frac{\arctan(k_{0,\alpha}\xi_0/2)}{4\pi k_{0,\alpha}}\theta(\Lambda-k_{0,\alpha}),
\end{equation}
one derives that the correction (\ref{eqn_corr_to_xi0}) to $\xi_0^{-2}$ contains an additional factor $(M+1)/L$ as compared to the standard case ($L=1$, $M=0$). Using large-$N$ expansion of the index $\nu$ for the standard case (see, e.g., \cite{Ma_1976}), we finally get the result $\nu=1-\frac{M+1}{L}\frac{32}{3\pi^2N}$ (see the third column of the table \ref{table_2}). For all the classes the factor $\frac{M+1}{L}$ is identical to previously introduced factor $r$.
\begin{table*}[t]      
\caption{\label{table_2} Critical indices to the first order in $1/N$}
\begin{tabular}{|c|c|c|}
\hline
classes                   & \hspace{1cm} $\eta$ \hspace{1cm} & \hspace{1cm} $\nu$ \hspace{1cm} \\
\hline
\hline
ferromagnet, A, A$^{\prime}$, A$_4$           &  $\frac{8}{3\pi^2 N}$   &  $1-\frac{32}{3\pi^2 N}$ \\
\hline
B, B$^{\prime}$, B$^{\prime\prime}$, B$_2$    &  $\frac{32}{9\pi^2 N}$  &  $1-\frac{128}{9\pi^2 N}$ \\ 
\hline
\end{tabular}
\end{table*}

{\it In conclusion}, we have found two different types A and B of the magnetic structures each of which are highly degenerate at the mean-field level for $L>1$ (see table \ref{table_1}). For these classes the fluctuations lift the degeneracy in favour of the collinear order opening exchange gaps in the spin-wave spectrum, as well as in $\chi_t(\mathbf{k})^{-1}$. The exchange ``gaps'' in the latter are closing at approaching $T_\mathrm{N}$ making the critical behavior for the B classes non-conventional (see table \ref{table_2}). The study for the class A$_4$ (AF-II structure on fcc lattice in the notations of Ref.\cite{Yamamoto_JPSJ_1972}) for special choices of $J_{ij}$ was performed previously in Ref. \cite{Yildirim_fcc_PRB_1998} within spin wave expansion.

{\it The research was carried out within the state assignment of FASO of Russia (theme ``Quant'' No. 01201463332), supported in part by RFBR (project No. 16-32-00482).}

\bibliography{bibliography}


\end{document}